\documentstyle[prl,aps,epsf]{revtex}
\begin{document}

\title{ $SU(2)$-spin Invariant Auxiliary Field Quantum Monte-Carlo Algorithm 
for Hubbard models }

\author{F.F. Assaad}
\address{
   Institut f\"ur Theoretische Physik III, \\
   Universit\"at Stuttgart, Pfaffenwaldring 57, D-70550 Stuttgart, Germany. }

\maketitle
\begin{abstract}
Auxiliary field quantum Monte Carlo  methods for Hubbard models are
generally based on a Hubbard-Stratonovitch transformation  where the 
field couples to the $z$-component of the spin. 
This transformation breaks
$SU(2)$ spin invariance. The symmetry is restored only after summation over
the auxiliary fields. Here, we analyze an alternative decomposition, which
conserves $SU(2)$ spin invariance, but requires the use of complex numbers.
We show that this algorithm gets rid of the very large fluctuations observed
in imaginary time displaced 
correlation functions of quantities  which  do not commute with 
the $z$-component of the total spin. 
The algorithm prooves to be efficient for the study of spin dynamics.
\end{abstract}

Auxiliary field quantum Monte Carlo Algorithms for Hubbard models are 
usually based  on the discrete Hubbard-Stratonovtich decomposition 
\cite{Hirsch83}:
\begin{equation}
\label{HS1}
 \exp \left( - \Delta \tau  U \sum_{\vec{i}} 
\left( n_{\vec{i},\uparrow } - \frac{1}{2} \right)
\left( n_{\vec{i},\downarrow } - \frac{1}{2} \right) \right) 
 = \tilde{C} \sum_{s_1, \dots, s_N  = \pm 1 }
  \exp \left(  \tilde{\alpha} \sum_{\vec{i}}  s_{\vec{i}}
\left( n_{\vec{i},\uparrow}  -n_{\vec{i},\downarrow } \right) \right).
\nonumber
\end{equation}
Here, 
$ n_{\vec{i},\sigma} = c^{\dagger}_{\vec{i},\sigma}  c_{\vec{i},\sigma}$
where $ c^{\dagger}_{\vec{i},\sigma}$  ($c_{\vec{i},\sigma}$) creates
(annihilates) an electron on site $ \vec{i} $ with $z$-component of spin
$\sigma$, $\cosh(\tilde {\alpha})   = \exp \left( \Delta \tau U / 2  \right)$
On an $N$-site lattice,  the constant $\tilde{C}  = 
\exp\left( \Delta  \tau U N / 4 \right )/2^N $ and  
$ \Delta \tau $ corresponds to an imaginary time step.
As apparent from the above equation, for a fixed set of Hubbard-
Stratonovitch  (HS) fields, $ s_1 \dots s_N$,  $SU(2)$-spin symmetry 
is broken. (i.e. the expression is not invariant under the transformation
$c_{\vec{j},\sigma} \rightarrow  
\left[ \exp \left( i \phi \vec{e} \vec{\sigma}/2 \right) \right]_{\sigma,s'}
c_{\vec{j},s'}  $. Here, $\vec{e}$ is a unit vector and $\vec{\sigma}$ is a
vector  consisting of the Pauli-spin matrices.)
Clearly $SU(2)$ spin symmetry is restored after
summation over the HS fields. 

Alternatively,   one may consider \cite{Hirsch83}
\begin{equation}
\label{HS2}
 \exp \left( - \Delta \tau  U \sum_{\vec{i}} 
\left( n_{\vec{i},\uparrow } - \frac{1}{2} \right)
\left( n_{\vec{i},\downarrow } - \frac{1}{2} \right) \right) 
 =  C \sum_{s_1, \dots, s_N  = \pm 1 } 
   \exp \left( i  \alpha \sum_{\vec{i}}  s_{\vec{i}}
\left( n_{\vec{i},\uparrow} + n_{\vec{i},\downarrow }  - 1 \right) \right).
\nonumber
\end{equation}
where  $ \cos(\alpha)   = \exp \left( - \Delta \tau U / 2  \right)$
and $C  = \exp\left( \Delta  \tau U N / 4 \right )/2^N $.
With this choice of the HS  transformation 
$SU(2)$ spin invariance is retained for any given HS configuration.

The aim of this note, is to address the question: will we obtain
a more efficient and/or reliable quantum Monte Carlo  algorithm if we
enhance the number of symmetries conserved by the HS transformation.
We consider the extended Hubbard model:
\begin{equation}
\label{tUW}
      H =  -\frac{t}{2} \sum_{\vec{i}} K_{\vec{i}} 
	  - W \sum_{\vec{i}} K_{\vec{i}}^{2} + 
          U \sum_{\vec{i}}
         (n_{\vec{i},\uparrow}-\frac{1}{2})
         (n_{\vec{i},\downarrow} -\frac{1}{2})
\end{equation}
with the hopping kinetic energy
\begin{equation}
        K_{\vec{i}} = \sum_{\sigma, \vec{\delta}}
   \left(c_{\vec{i},\sigma}^{\dagger} c_{\vec{i} + \vec{\delta},\sigma} +
        c_{\vec{i} + \vec{\delta},\sigma}^{\dagger} c_{\vec{i},\sigma} \right).
\end{equation}
Here $W \geq 0$,  $\vec{\delta} = \pm \vec{a}_x, \pm \vec{a}_y $ where
$\vec{a}_x$, $\vec{a}_y$  are the lattice constants. 
We impose twisted boundary conditions:
\begin{equation}
\label{Bound}
 c_{\vec{i} + L \vec{a}_x, \sigma } = \exp \left(2 \pi i \Phi/\Phi_0
\right) c_{\vec{i}, \sigma}, \; \;  c_{\vec{i} + L \vec{a}_y, \sigma }
= c_{\vec{i}, \sigma},
\end{equation}
with $\Phi_0 = h c / e$ the flux quanta and  $L$ the  linear length
of the square lattice.
The boundary conditions given by Eq. (\ref{Bound})
account for a magnetic flux threading a torus on which  the lattice is
wrapped. 
At half-filling, and 
constant value of $U/t$  the $W$-term drives the  ground state from a
antiferromagnetic Mott insulator to a $d_{x^2 - y^2}$  superconductor
\cite{Assaad_tUW_1}. At $U/t=4$, this quantum transition occurs 
at $W_c/t \sim 0.3 $.
At finite values of  $W < W_c$, numerical simulations are consistent with the
occurrence of a  $d_{x^2 - y^2}$ superconductor upon doping of the 
Mott insulating state \cite{Assaad_tUW_2}.

To decompose the $W$-term, we use the approximate relation
\begin{equation}
\label{HSW}
	e^{\Delta \tau W K_{\vec{i}}^{2}  } =   \frac{1}{4}
    \sum_{l = -2, -1, 1, 2 } \gamma (l) 
\exp  \left(  \sqrt{ \Delta \tau W } \eta (l) K_{\vec{i}} \right) + 
O(\Delta \tau ^4),
\end{equation}
where the fields $\eta$ and $\gamma$ take the values:
\begin{eqnarray}
 \gamma(\pm 1) = 1 + \sqrt{6}/3, \; \; \gamma(\pm 2) = 1 - \sqrt{6}/3
\nonumber \\
 \eta(\pm 1 ) = \pm \sqrt{2 \left(3 - \sqrt{6} \right)},  \; \;
 \eta(\pm 2 ) = \pm \sqrt{2 \left(3 + \sqrt{6} \right)} 
\nonumber  \\
\end{eqnarray}
Since $K_{\vec{i}} $ is invariant under  a rotation in spin space, the above
Hubbard-Stratonovitch decomposition conserves $SU(2)$-spin symmetry. 
Thus,  the choice of the HS transformation for the Hubbard  term
will determine
whether the algorithm is $SU(2)$-spin invariant  or not. 

We have carried out our simulations with the Projector QMC algorithm
\cite{Koonin,Sandro}.
Within this approach,  the ground state expectation value of an observable $O$ 
is obtained with:
\begin{equation}
\frac{\langle \Psi_0 | O |  \Psi_0 \rangle }
           {\langle \Psi_0 |  \Psi_0 \rangle }
           = \lim_{ \Theta \rightarrow \infty }
  \frac{ \langle \Psi_T |e^{-\Theta H }
          O
         e^{-\Theta H } | \Psi_T \rangle }
       { \langle \Psi_T |e^{-2\Theta H } | \Psi_T \rangle }.
\end{equation}
The  ground state $ |  \Psi_0 \rangle $ is filtered out of a trial wave
function,  $ | \Psi_T \rangle $, provided that
$ \langle  \Psi_0  | \Psi_T \rangle  \neq 0 $.
We choose the trial wave function to be a spin singlet solution
of the non interacting Hamiltonian (U=W=0). An explicit construction
of such trial wave functions may be found in reference  \cite{Assaad_tUW_1}.

After Trotter decomposition of the imaginary time propagation and 
HS  transformation of  the two-body terms,  one obtains:
\begin{equation}
\frac{ \langle \Psi_T |e^{-\Theta H }
          O
         e^{-\Theta H } | \Psi_T \rangle }
       { \langle \Psi_T |e^{-2\Theta H } | \Psi_T \rangle } 
= \sum_{x} {\rm Pr } (\Theta, x )  \langle O \rangle (\Theta,x)
 + 
O(\Delta \tau ^2),
\end{equation} 
where  $x$ denotes a configuration of HS fields,
and $ \langle O \rangle (\Theta,x) $, corresponds to the value of the 
observable $O$  the the HS fields $x$.
At half-band filling particle-hole symmetry leads to  positive  
values of 
${\rm Pr } (\Theta, x ) $ which may thus be interpreted as a probability
distribution and sampled with Monte Carlo methods.  This statement is
valid for both choices of the HS transformation of the Hubbard term 
(\ref{HS1},  \ref{HS2}).

We now compare the $SU(2)$ spin invariant algorithm based  on
Eq. (\ref{HS2})  to the 
$SU(2)$ spin non-invariant algorithm  based on Eq. (\ref{HS1}).
The $SU(2)$ spin-invariant algorithm forces us to work with
complex numbers. On the other hand, for many applications real numbers
may be used for the $SU(2)$ non-invariant code. For the comparison  
discussed below, and to keep the CPU time approximately constant,
we have carried out  twice as many sweeps for the 
$SU(2)$ non-invariant code than for the $SU(2)$ invariant code.
We consider various observables at half-band filling.

a) {\bf Magnetization}.  In the case  of the $SU(2)$-invariant algorithm,
and the above mentioned choices of the trial wave function one has:
\begin{equation}
\label{GRSU2}
  \langle c_{\vec{i},\uparrow}^{\dagger} 
   c_{\vec{j}, \uparrow}  \rangle (\Theta,x) = 
  \langle c_{\vec{i},\downarrow}^{\dagger} c_{\vec{j}, \downarrow}  
  \rangle (\Theta,x). 
\end{equation}
Thus,  the total magnetization, 
$   m_z(\vec{q}) = \sum_{\vec{j}} e^{i \vec{q} \vec{j} } 
	\left( n_{i,\uparrow} - n_{i,\downarrow} \right) $
is identical to zero for all  values of the HS fields:
\begin{equation}
\langle m_z(\vec{q})  \rangle (\Theta, x )  \equiv 0.
\end{equation}
On the other hand, the $SU(2)$ non-invariant algorithm,  equation
(\ref{GRSU2}) is not valid, and one obtains  zero magnetization, only
after summation over the HS fields.   At $\vec{q} = (\pi,\pi) \equiv \vec{Q} $,
$L=6$, $\langle n \rangle = 1$, $U/t = 4$ and $W/t = 0$ one obtains after
$2\times 10^5$ sweeps,  $ \langle m_z(\vec{Q})  \rangle = -0.16 \pm 0.31$. 
Thus for this trivial case, the advantage
of the $SU(2)$ invariant algorithm  over the $SU(2)$ 
non-invariant algorithm is infinite.

\begin{figure}[ht]
\epsfxsize=8cm
\epsfxsize=10cm
\hfil\epsfbox{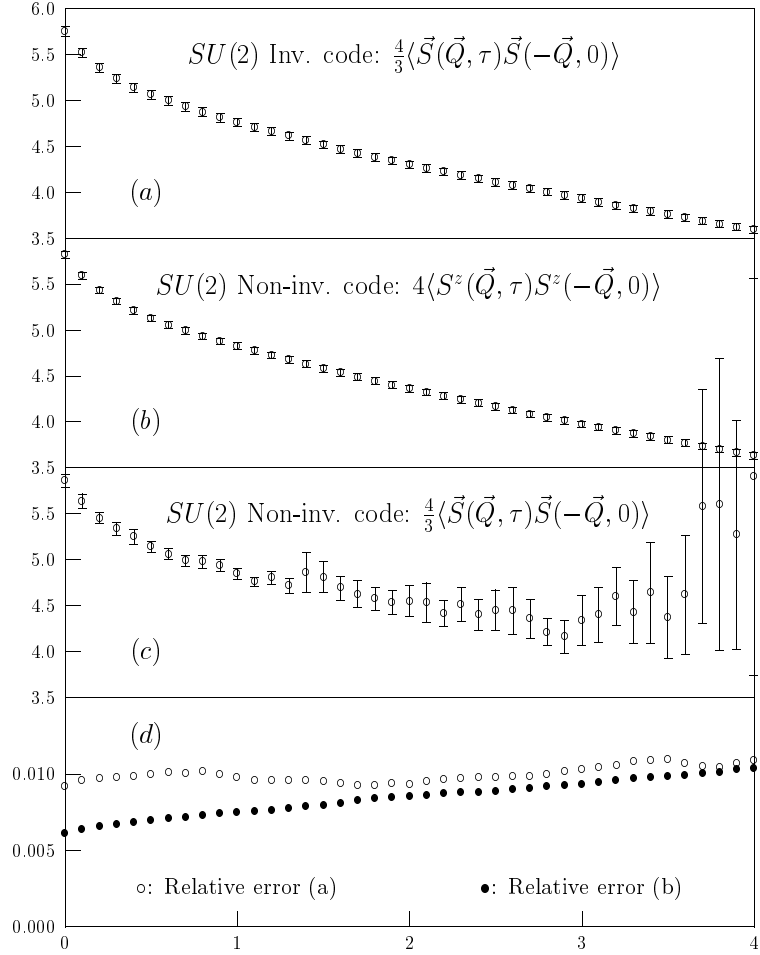}\hfil
\caption[]
{\noindent  Imaginary time displaced spin-spin correlations, at
$\vec{Q} = (\pi,\pi)$. The $SU(2)$ invariant code is based on the
HS transformation of Eq.  (\ref{HS2}) and  $SU(2)$  non-invariant
on Eq.  (\ref{HS1}). To keep the CPU time approximately constant between the
two simulations, we have carried out twice as many sweeps for
the $SU(2)$ non-invariant code as for the $SU(2)$ invariant code.
Here, we have used periodic boundary conditions, $\Phi=0$ in Eqn.
(\ref{Bound}).
\label{tU.fig} }
\end{figure}

b) {\bf Imaginary time displaced spin-spin correlations}.  Here  
we consider the quantity: 
$ S^{\alpha} (\vec{q},\tau) S^{\alpha} (-\vec{q},0) $ where
$ S^{\alpha} (\vec{q},\tau)  = 
\sum_{\vec{j}} e^{i \vec{q} \vec{j}}  e^{\tau H} S^{\alpha}_{j} e^{-\tau H} $,
$S^{\alpha}_{j}$ being the  $\alpha$-component of the spin operator
on site $\vec{j}$.   The numerically stable computation of imaginary
time displaced correlation functions within the Projector QMC algorithm 
is described in Ref. \cite{Assaad_GRT0}. 
In the SU(2)-invariant algorithm one obtains with the above mentioned trial
wave function:
\begin{equation}
 \langle S^{\alpha} (\vec{q},\tau) S^{\alpha} (-\vec{q},0) \rangle (\Theta, x ) \equiv
 \langle S^{\gamma} (\vec{q},\tau) S^{\gamma} (-\vec{q},0) \rangle (\Theta, x ). 
\end{equation}
Here, $\alpha, \gamma$ run over the three components of the spin. 
In the case of the SU(2) non-invariant  the above equation is valid
only after summation over the  HS fields, $x$. 
Fig. 1 plots the spin-spin correlations for
$\langle n \rangle = 1$, $U/t = 4$ and $W/t = 0$ on a $6 \times 6$ lattice.
The half-filled Hubbard model is expected to show long-range antiferromagnetic 
order in the thermodynamic limit.  Thus,  
$ \frac{1}{N} \langle  \vec{S} (\vec{Q},\tau) \vec{S} (-\vec{Q},0) \rangle $ is 
should  saturate to
a finite value in the {\it large}  $L$ and $\tau$  limits. Here, $L$ is the 
linear size of the square lattice and $N$ the number of sites.  
On a finite size lattice, a spin gap is expected, thus leading to an 
exponential decay 
in $\tau$ of the considered quantity. 
If one compares the quantity 
$ \langle  \vec{S} (\vec{Q},\tau) \vec{S} (-\vec{Q},0) \rangle $ for both codes,
(Fig. 1a and Fig. 1c) it is clear that the $SU(2)$ invariant code does 
much better.  
The  large fluctuations in the case of the $SU(2)$ non-invariant code
may be traced back to the $x$ and $y$-components of the spin-spin correlation
function. In fact, considering only the $z$-component of the correlation 
function (Fig. 1b) yields good results. The HS transformation of
equation (\ref{HS1}) conserves the $z$-component of the total spin, but not
the other components. 
We now consider the $z$-component of the spin-spin correlations
and compare both algorithms. We have plotted 
in Fig. 1d  the relative errors for both codes. 
\begin{figure}[ht]
\epsfxsize=8cm
\epsfxsize=10cm
\hfil\epsfbox{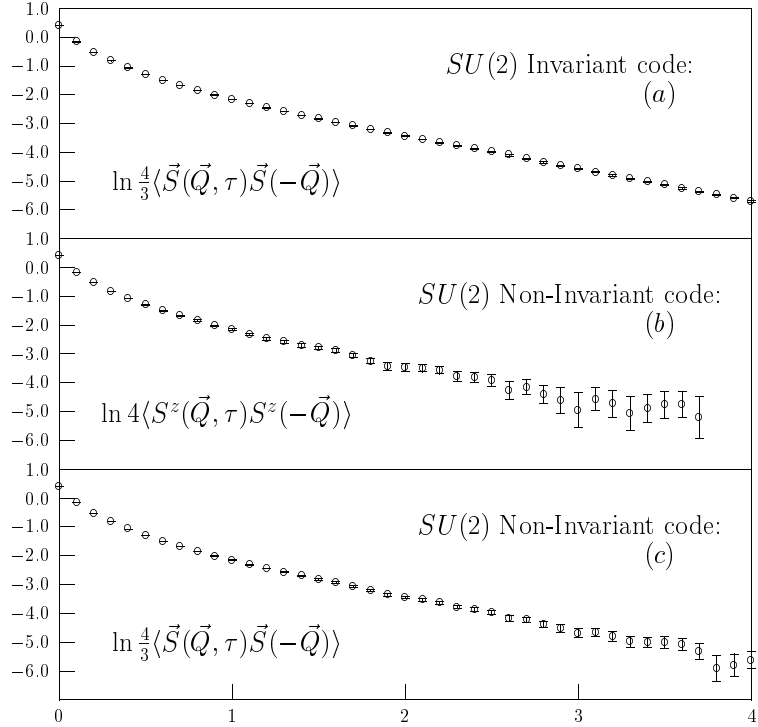}\hfil
\caption[]
{\noindent Same as Fig. \ref{tU.fig} for the parameter set:
$W/t = 0.35$, $U/t = 2$ and at half-band filling.  Here we
use antiperiodic boundary conditions set $\Phi = \Phi_0/2$.
It is clear that the $SU(2)$-invariant code does substantially
better at {\it large} values of  $\tau$.
\label{tUW1.fig} }
\end{figure}
Overall, the SU(2) 
spin-invariant  code  produces larger errors within the  considered  
$\tau$-range. However, for the $SU(2)$-invariant code the relative error, is
to a first approximation independent of $\tau$. In contrast, 
the relative error for the $SU(2)$ non-invariant code grows  as a function
of $\tau$.  
At large values of $\tau$ we expect the  the $SU(2)$ invariant code
to be more efficient. 
To confirm this statement, we consider the
parameter set: $\langle n \rangle = 1$, 
$U/t = 2$, $W/t = 0.35 $, $\Phi= \Phi_0/2$  on a $8 \times 8$ lattice. 
The data is shown in Fig.2.  It is clear that the $SU(2)$-invariant code
does much better at {\it large} values of $\tau t$. In Fig. 2, the
$z$-component of the spin-spin correlations ( Fig. 2b) shows 
larger fluctuations than  
$ \langle  \vec{S} (\vec{Q},\tau) \vec{S} (-\vec{Q},0) \rangle $ 
( Fig. 2c). Exactly the opposite is seen in Figs. 1b and 1c. 

\begin{table}
\begin{center}
\begin{tabular}{|c|c|c|c|} 
\hline
$ \tau t$ & $ SU(2)$ invariant code  &
$ SU(2)$ non-invariant code   &
$ SU(2)$ non-invariant code  \\
   &
$ \langle \Pi^{z}(\tau) \Pi^{z,\dagger}(0)\rangle $ & 
$ \langle \Pi^{x}(\tau) \Pi^{x,\dagger}(0)\rangle $ & 
$ \langle \Pi^{z}(\tau) \Pi^{z,\dagger}(0)\rangle $ \\ 
\hline
$0  $& $ 0.15055 \pm 0.00080 $ & $ 0.15097 \pm 0.00073 $ & $ 0.13837 \pm 0.01506 $ \\
$0.2$& $ 0.05902 \pm 0.00058 $ & $ 0.05920 \pm 0.00051 $ & $ 0.05845 \pm 0.00387 $ \\
$0.4$& $ 0.02915 \pm 0.00055 $ & $ 0.02941 \pm 0.00040 $ & $ 0.02846 \pm 0.00215 $ \\
$0.6$& $ 0.01474 \pm 0.00046 $ & $ 0.01547 \pm 0.00032 $ & $ 0.01373 \pm 0.00254 $ \\
$0.8$& $ 0.00728 \pm 0.00046 $ & $ 0.00859 \pm 0.00039 $ & $ 0.00841 \pm 0.00105 $ \\
$1  $& $ 0.00397 \pm 0.00038 $ & $ 0.00480 \pm 0.00053 $ & $ 0.00394 \pm 0.00104 $ \\
$1.2$& $ 0.00240 \pm 0.00033 $ & $ 0.00251 \pm 0.00033 $ & $ 0.00086 \pm 0.00080 $ \\
\hline
\end{tabular}
\end{center}
\caption{$\Pi$-mode correlation functions  at half-filling 
for $U/t = 4$, $W/t = 0.35 $ 
on a $6 \times 6$
lattice. Data in the last two columns were obtained with the SU(2) non-invariant
code. Within this algorithm, the $x$ and $y$ components of the $\Pi$-mode
correlation  function are identical. For the $SU(2)$ spin-invariant 
algorithm, the results are independent on the considered  components of the 
correlation function.
To keep the CPU time approximately constant between the
two simulations, we have carried out twice as many sweeps for
the $SU(2)$ non-invariant code as for the $SU(2)$ invariant code. }
\end{table}

c) {\bf $\Pi$-modes}. The $\Pi$ modes  introduced in the $SO(5)$ 
theory of the unification of antiferromagnetism and superconductivity
are defined by: $ \Pi^{\alpha}  =  \sum_{\vec{p}, s, s' } 
g(\vec{p}) c_{\vec{p} + \vec{Q}, s }
 \left( \sigma^{\alpha} \sigma^{y} \right)_{s,s'}  
c_{\vec{p}, s'} $ \cite{Zhang}. 
Here, $\sigma^{\alpha}$ corresponds to the Pauli spin matrix
and for the case of $d$-wave superconductivity,  we consider $g(\vec{p})
= \cos(p_x) - cos(p_y) $. The  $SU(2)$-spin invariant code
satisfies
\begin{equation}
 \langle \Pi^{\alpha}(\tau) \Pi^{\alpha,\dagger}(0)\rangle (\Theta, x ) \equiv
 \langle \Pi^{\gamma}(\tau) \Pi^{\gamma,\dagger}(0)\rangle (\Theta, x ). 
\end{equation}
In the case of the $SU(2)$ non-invariant code,  the above equation is valid
for all values of $\gamma$ and $\alpha$ only after summation over 
the HS fields $x$.
In the table,  imaginary time $\Pi$ correlations are considered   for
both algorithms. 
For the $SU(2)$ non-invariant code, substantial fluctuations
in  $\langle \Pi^{z}(\tau) \Pi^{z,\dagger}(0)\rangle $ are observed at 
{\it large } values of $\tau$.  On the other hand, the error-bars in 
 $\langle \Pi^{x}(\tau) \Pi^{x,\dagger}(0)\rangle $  are smaller. We again
attribute the large fluctuations in the $z$ component of the $\Pi$ mode
correlations to fact that $\Pi^{z}$ does not commute with the $z$-component
of the total spin. In contrast,  $\Pi^{x}$  does commute with the 
$z$-component of the total spin.   The $SU(2)$-invariant code shows good
convergence, and the results agree with those obtained for the
x-component of the $\Pi$-mode within the $SU(2)$ non-invariant algorithm.

d) {\bf Single particle Green functions.}
We have not found any significant improvements in the fluctuations of the 
imaginary time single particle Green functions between the two algorithms.
To be more precise, the fluctuation of 
$ \frac{1}{N} \sum_{ \sigma , \vec{i} } \langle c_{\vec{i} ,\sigma }
(\tau) c^{\dagger}_{\vec{i} ,\sigma} \rangle  $
are to a first approximation
independent  of the choice of HS transformation.  Thus for both considered 
codes, the error-bars scale as $C/\sqrt{N_{sweeps}}$  where $C$ is a 
constant  independent on the choice of the algorithm, and  $N_{sweeps}$ denotes
the number of sweeps carried out.

In conclusion, we have considered an alternative HS transformation for
Hubbard type models which conserves $SU(2)$ spin symmetry.
This algorithm requires the use of complex numbers and does not introduce
a sign problem at half-band filling. In the case of $SU(2)$ non-invariant 
algorithms based on Eq. (\ref{HS1}), very large fluctuations  can 
occur in  the calculation  of imaginary time displaced  quantities of the form
 $ \langle  A(\tau) A^{\dagger}  \rangle $ when $A$ does not commute with
the $z$-component of the  total spin.  In the $SU(2)$ invariant formulation,
this pathology does not occur.
We have compared the two algorithms  in the worst case scenario where
real numbers can be used  for  the $SU(2)$  non-invariant 
algorithm.  By keeping the
CPU time constant for both codes we have shown that the 
long imaginary time behavior of the spin-spin correlations 
are obtained  more efficiently with
the $SU(2)$ invariant code than with the  $SU(2)$ non-invariant code.
Thus, the $SU(2)$ invariant code is  more efficient  for the measure
of spin gaps which may be extracted from the long imaginary time
decay of the spin-spin correlations.  More generally,  it is an efficient
code for the study of spin dynamics.
The fluctuations of other quantities such as  single particle green function, 
were invariant under the choice of the HS transformation.

\section*{Acknowledgments}
M. Muramatsu is thanked for many instructive conversations.
The computations were carried out on the T3E of the HLRS,Stuttgart, as
well as on the T90  and T3E of the HLRZ, J\"ulich.

\end{document}